\documentclass[letterpaper, 10 pt, conference, dvipsnames]{ieeeconf} 
\IEEEoverridecommandlockouts
\newcommand\copyrighttext{%
	\footnotesize This work has been submitted to the IEEE for possible publication. Copyright may be transferred without notice, after which this version may no longer be accessible.%
}
\newcommand\copyrightnotice{%
	\begin{tikzpicture}[remember picture,overlay]
	\node[anchor=south,yshift=10pt, xshift=10pt] at (current page.south) {\fbox{\parbox{\dimexpr\textwidth-\fboxsep-\fboxrule\relax}{\copyrighttext}}};
	\end{tikzpicture}%
}

\usepackage{printlen}

\usepackage[T1]{fontenc}
\usepackage{amsmath,amsfonts, amssymb}
\usepackage{cite}
\usepackage{stfloats}
\usepackage{graphicx}
\usepackage[caption=false,font=scriptsize]{subfig}
\usepackage{url}
\usepackage[hidelinks]{hyperref}

\usepackage[per-mode = power]{siunitx}
\usepackage{acronym}
\usepackage[none]{hyphenat}
\usepackage{glossaries}
\glsdisablehyper
\usepackage[switch]{lineno}

\usepackage{algorithm}
\usepackage{algpseudocode}

\usepackage{booktabs}
\usepackage{tabularx}
\usepackage{multirow}
\usepackage{pifont}

\usepackage{wheelchart}
\usepackage{etoolbox}
\usepackage{tikzsymbols}

\usepackage{tikz, pgfplots, pgfplotstable}
\usetikzlibrary{positioning, arrows.meta, fit, backgrounds, calc, shapes.geometric, decorations.markings, patterns, decorations.pathreplacing, fadings, shadows, decorations.pathreplacing, calligraphy}
\usepgfplotslibrary{statistics}
\usepgfplotslibrary{groupplots}
\usepgfplotslibrary{fillbetween}
\pgfplotsset{compat=1.18}

\usepackage{xcolor}
\definecolor{Black}{HTML}{000000}
\definecolor{Blue}{HTML}{0065bd}
\definecolor{Bluelight}{HTML}{D6E8F7}

\definecolor{Bluestrong}{HTML}{003359}

\definecolor{OrangePP}{HTML}{E97132}
\definecolor{Green}{HTML}{A2AD00}
\definecolor{GreenCR}{HTML}{008000}
\definecolor{LightGray}{HTML}{e7e7e7}
\definecolor{Gray}{HTML}{7f7f7f}
\definecolor{Gray-opac}{HTML}{d8d8d8}
\definecolor{MyDarkBlue}{RGB}{14,40,65}
\definecolor{Red}{RGB}{196,7,27}

    \definecolor{TUMBlue}{HTML}{0065BD}
    \definecolor{TUMBlack}{HTML}{000000}
    \definecolor{TUMWhite}{HTML}{FFFFFF}
    \definecolor{TUMDarkBlue}{HTML}{005293}
    \definecolor{TUMLightBlue}{HTML}{64A0C8}
    \definecolor{TUMLighterBlue}{HTML}{98C6EA}
    \definecolor{TUMGray}{HTML}{999999}
    \definecolor{TUMOrange}{HTML}{E37222}
    \definecolor{TUMGreen}{HTML}{A2AD00}
    \definecolor{TUMLightGray}{HTML}{DAD7CB}

    \definecolor{TUMBlueBrand}{HTML}{3070B3}
    \definecolor{TUMBlueDark}{HTML}{072140}
    \definecolor{TUMBlueDark1}{HTML}{0A2D57}
    \definecolor{TUMBlueDark2}{HTML}{0E396E}
    \definecolor{TUMBlueDark3}{HTML}{114584}
    \definecolor{TUMBlueDark4}{HTML}{14519A}
    \definecolor{TUMBlueDark5}{HTML}{165DB1}
    \definecolor{TUMBlueLight}{HTML}{5E94D4}
    \definecolor{TUMBlueLightDark}{HTML}{9ABCE4}
    \definecolor{TUMBlueLight2}{HTML}{C2D7EF}
    \definecolor{TUMBlueLight3}{HTML}{D7E4F4}
    \definecolor{TUMBlueLight4}{HTML}{E3EEFA}
    \definecolor{TUMBlueLight5}{HTML}{F0F5FA}
    
    \definecolor{TUMYellow}{HTML}{FED702}
    \definecolor{TUMYellowDark}{HTML}{CBAB01}
    \definecolor{TUMYellow1}{HTML}{FEDE34}
    \definecolor{TUMYellow2}{HTML}{FEE667}
    \definecolor{TUMYellow3}{HTML}{FEEE9A}
    \definecolor{TUMYellow4}{HTML}{FEF6CD}
    
    \definecolor{TUMWebOrange}{HTML}{F7811E}
    \definecolor{TUMOrangeDark}{HTML}{D99208}
    \definecolor{TUMOrange1}{HTML}{F9BF4E}
    \definecolor{TUMOrange2}{HTML}{FAD080}
    \definecolor{TUMOrange3}{HTML}{FCE2B0}
    \definecolor{TUMOrange4}{HTML}{FEF4E1}
    
    \definecolor{TUMPink}{HTML}{B55CA5}
    \definecolor{TUMPinkDark}{HTML}{9B468D}
    \definecolor{TUMPink1}{HTML}{C680BB}
    \definecolor{TUMPink2}{HTML}{D6A4CE}
    \definecolor{TUMPink3}{HTML}{E6C7E1}
    \definecolor{TUMPink4}{HTML}{F6EAF4}
    
    \definecolor{TUMBlueBright}{HTML}{8F81EA}
    \definecolor{TUMBlueBrightDark}{HTML}{6955E2}
    \definecolor{TUMBlueBright1}{HTML}{B6ACF1}
    \definecolor{TUMBlueBright2}{HTML}{C9C2F5}
    \definecolor{TUMBlueBright3}{HTML}{DCD8F9}
    \definecolor{TUMBlueBright4}{HTML}{EFEDFC}
    
    \definecolor{TUMRed}{HTML}{EA7237}
    \definecolor{TUMRedDark}{HTML}{D95117}
    \definecolor{TUMRed1}{HTML}{EF9067}
    \definecolor{TUMRed2}{HTML}{F3B295}
    \definecolor{TUMRed3}{HTML}{F6C2AC}
    \definecolor{TUMRed4}{HTML}{FBEADA}
    
    \definecolor{TUMWebGreen}{HTML}{9FBA36}
    \definecolor{TUMGreenDark}{HTML}{7D922A}
    \definecolor{TUMGreen1}{HTML}{B6CE55}
    \definecolor{TUMGreen2}{HTML}{C7D97D}
    \definecolor{TUMGreen3}{HTML}{D8E5A4}
    \definecolor{TUMGreen4}{HTML}{E9F1CB}
    
    \definecolor{TUMGrey1}{HTML}{20252A}
    \definecolor{TUMGrey2}{HTML}{333A41}
    \definecolor{TUMGrey3}{HTML}{475058}
    \definecolor{TUMGrey4}{HTML}{6A757E}
    \definecolor{TUMGrey7}{HTML}{DDE2E6}
    \definecolor{TUMGrey8}{HTML}{EBECEF}
    \definecolor{TUMGrey9}{HTML}{FBF9FA}
    \definecolor{TUMWebWhite}{HTML}{FFFFFF}

\graphicspath{{./figures/}}

\title{\LARGE \bf
Spatio-Temporal Calibration of a Vehicle-Mounted Microphone Array for Autonomous Driving}

\author{Marvin Seegert, Richard Taupitz, Korbinian Moller, Johannes Betz%
    \thanks{M. Seegert, K. Moller and J. Betz are with the Professorship of Autonomous Vehicle Systems and R. Taupitz with the Institute of Automotive Technology, TUM School of Engineering and Design, Technical University of Munich, 85748 Garching, Germany; Munich Institute of Robotics and Machine Intelligence (MIRMI).}%
    }

\newacronym{av}{AV}{Autonomous Vehicle}%
\newacronym{cfd}{CFD}{Constant Fraction Discriminator}
\newacronym{tdoa}{TDOA}{Time Difference of Arrival}
\newacronym{ransac}{RANSAC}{Random Sample Consensus}
\newacronym{roi}{ROI}{Region of Interest}
\newacronym{rms}{RMS}{Root Mean Square}
\newacronym{los}{LOS}{Line-of-Sight}
\newacronym{nlos}{NLOS}{Non-Line-of-Sight}
\newacronym{cad}{CAD}{Computer-Aided Design}
\newacronym{snr}{SNR}{Signal-to-Noise Ratio}
\newacronym{lidar}{LiDAR}{Light Detection and Ranging}
\newacronym{gcc}{GCC}{Generalized Cross-Correlation}
\newacronym{vrtf}{VRTF}{Vehicle-Related Transfer Function}
\newacronym{ptp}{PTP}{Precision Time Protocol}
\newacronym{slam}{SLAM}{Simultaneous Localization and Mapping}

\newcommand{\ablBlockedRays}{319}
\newcommand{\ablClearRays}{321}
\newcommand{\ablClosureSigmaCm}{6.06}
\newcommand{\ablClosureSigmaMs}{0.17}
\newcommand{\ablClosureTriangles}{4928}
\newcommand{\ablDistMaskInLosFull}{6.76}
\newcommand{\ablDistMaskInLosRef}{7.09}
\newcommand{\ablFoldHi}{7.55}
\newcommand{\ablFoldLo}{4.84}
\newcommand{\ablFolds}{5}
\newcommand{\ablFreeCDelta}{-0.75}
\newcommand{\ablFreeCHeldout}{5.62}
\newcommand{\ablFreeCInLos}{5.50}
\newcommand{\ablFreeCInNlos}{17.08}
\newcommand{\ablFreeCOutliers}{1}
\newcommand{\ablFreeCSE}{0.14}
\newcommand{\ablFreeCShift}{43.36}
\newcommand{\ablFreeCSpeed}{288.9}
\newcommand{\ablFreeCTemp}{-70}
\newcommand{\ablMaskAgreement}{81.4}
\newcommand{\ablMaskFullDelta}{+4.97}
\newcommand{\ablMaskFullHeldout}{11.34}
\newcommand{\ablMaskFullInLos}{14.15}
\newcommand{\ablMaskFullInNlos}{29.27}
\newcommand{\ablMaskFullOutliers}{28}
\newcommand{\ablMaskFullSE}{0.96}
\newcommand{\ablMaskFullShift}{23.36}
\newcommand{\ablMaskPartialDelta}{+0.75}
\newcommand{\ablMaskPartialHeldout}{7.12}
\newcommand{\ablMaskPartialInLos}{7.03}
\newcommand{\ablMaskPartialInNlos}{25.78}
\newcommand{\ablMaskPartialOutliers}{15}
\newcommand{\ablMaskPartialSE}{0.36}
\newcommand{\ablMaskPartialShift}{5.48}
\newcommand{\ablMeasuredCHi}{349.68}
\newcommand{\ablMeasuredCLo}{348.98}
\newcommand{\ablNlosExcess}{11.76}
\newcommand{\ablNoCovDelta}{+0.00}
\newcommand{\ablNoCovHeldout}{6.37}
\newcommand{\ablNoCovInLos}{6.35}
\newcommand{\ablNoCovInNlos}{25.08}
\newcommand{\ablNoCovOutliers}{13}
\newcommand{\ablNoCovSE}{0.01}
\newcommand{\ablNoCovShift}{0.43}
\newcommand{\ablNoSnrDelta}{+1.38}
\newcommand{\ablNoSnrHeldout}{7.75}
\newcommand{\ablNoSnrInLos}{8.49}
\newcommand{\ablNoSnrInNlos}{20.83}
\newcommand{\ablNoSnrOutliers}{8}
\newcommand{\ablNoSnrSE}{0.44}
\newcommand{\ablNoSnrShift}{8.14}
\newcommand{\ablNoWeightDelta}{+1.39}
\newcommand{\ablNoWeightHeldout}{7.76}
\newcommand{\ablNoWeightInLos}{8.50}
\newcommand{\ablNoWeightInNlos}{20.82}
\newcommand{\ablNoWeightOutliers}{8}
\newcommand{\ablNoWeightSE}{0.44}
\newcommand{\ablNoWeightShift}{8.25}
\newcommand{\ablOneSidedDelta}{+0.08}
\newcommand{\ablOneSidedHeldout}{6.44}
\newcommand{\ablOneSidedInLos}{6.95}
\newcommand{\ablOneSidedInNlos}{29.40}
\newcommand{\ablOneSidedOutliers}{22}
\newcommand{\ablOneSidedSE}{0.15}
\newcommand{\ablOneSidedShift}{5.79}
\newcommand{\ablPredictedFloor}{6.07}
\newcommand{\ablRaySigmaPos}{0.34}
\newcommand{\ablRefHeldout}{6.37}
\newcommand{\ablRefInLos}{6.37}
\newcommand{\ablRefInNlos}{25.14}
\newcommand{\ablRefOutliers}{13}
\newcommand{\ablRefShift}{0.00}
\newcommand{\ablSigmaPos}{0.34}
\newcommand{\ablTotalRays}{640}
\newcommand{\ablTrustedMics}{4}
\newcommand{\cntBalancedOneTwo}{7.92}
\newcommand{\cntDraws}{40}
\newcommand{\cntElevHi}{29}
\newcommand{\cntElevLo}{-6}
\newcommand{\cntElevMed}{10}
\newcommand{\cntEvalN}{16}
\newcommand{\cntEvenEight}{12.7}
\newcommand{\cntFloor}{5.67}
\newcommand{\cntHorizOneSix}{9.3}
\newcommand{\cntHorizThreeTwo}{4.8}
\newcommand{\cntPlaceHi}{32}
\newcommand{\cntPlaceLo}{8}
\newcommand{\cntPlaceMinCell}{9}
\newcommand{\cntPlaceRuns}{2120}
\newcommand{\cntRefN}{64}
\newcommand{\cntResOneSix}{7.45}
\newcommand{\cntResOneTwo}{10.47}
\newcommand{\cntResThreeTwo}{6.58}
\newcommand{\cntUnevenThreeTwo}{18.7}
\newcommand{\cntVertOneSix}{18.4}
\newcommand{\dgDraws}{80}
\newcommand{\dgJitterOne}{5.5}
\newcommand{\dgSingleOne}{3.0}
\newcommand{\dgTransTen}{2.3}
\newcommand{\pbAzHi}{38}
\newcommand{\pbAzLo}{-16}
\newcommand{\pbBands}{4}
\newcommand{\pbBias}{-2.3}
\newcommand{\pbBiasSE}{1.1}
\newcommand{\pbClosureFront}{99}
\newcommand{\pbClosureMixed}{31}
\newcommand{\pbDuration}{11.5}
\newcommand{\pbExtent}{9}
\newcommand{\pbGateDb}{10}
\newcommand{\pbMedErr}{1.9}
\newcommand{\pbN}{93}
\newcommand{\pbNeff}{14}
\newcommand{\pbQOne}{0.9}
\newcommand{\pbQThree}{4.9}
\newcommand{\pbRangeHi}{9.7}
\newcommand{\pbRangeLo}{6.1}
\newcommand{\pbWindowMs}{250}
\newcommand{\pbWithinTen}{98}
\newcommand{\tempBagDiffSigma}{0.2}
\newcommand{\tempDriftSigma}{0.6}
\newcommand{\tempEpsilon}{5}
\newcommand{\tempFrameMs}{50}
\newcommand{\tempFrameRate}{20}
\newcommand{\tempInflation}{0.8}
\newcommand{\tempNBags}{2}
\newcommand{\tempNBursts}{51}
\newcommand{\tempNCams}{5}
\newcommand{\tempNSolved}{80}
\newcommand{\tempOffset}{-4.22}
\newcommand{\tempOffsetHalfWidth}{0.98}
\newcommand{\tempSigmaC}{14.43}
\newcommand{\tempSigmaJitterBound}{1.8}
\newcommand{\tempSigmaMin}{2.1}
\newcommand{\tempSigmaTotal}{14.58}

\usepackage{microtype}

\begin{document}
\bstctlcite{BSTcontrol}

\maketitle
\copyrightnotice

\begin{abstract}
Microphones let an autonomous vehicle sense into areas its cameras and LiDAR cannot see, by picking up the sound that other traffic participants emit. Locating those sources and fusing them with the visual perception requires the microphone array to be calibrated and registered relative to the vehicle's other sensors. Existing calibration methods assume small, unobstructed indoor rigs, whereas a vehicle body casts an acoustic shadow that straight-line propagation cannot represent, and to our knowledge no calibration of an array distributed around a vehicle has been reported. We propose a pipeline that uses a balloon burst as a single cross-modal target, which the LiDAR locates in metric coordinates, every microphone hears as a sharp onset and a camera times. Each observation is weighted by its own timing variance, which discounts shadowed arrivals instead of discarding them and needs no model of the vehicle. On a full-size vehicle with eight microphones over a $\mathbf{4.9}$\,m aperture, the held-out range residual of $\mathbf{\ablRefHeldout}$\,cm approaches the $\mathbf{\ablPredictedFloor}$\,cm floor set by the onset timing noise, and the calibrated array tracks a moving source to within a few degrees of the vehicle's own LiDAR. A dozen bursts spread around the vehicle suffice, and spreading them matters more than adding more.
\end{abstract}

\makeatletter
\def\abl@stripplus#1{\if+#1\else#1\fi}
\newcommand{\unsigned}[1]{\expandafter\abl@stripplus#1}
\makeatother

\section{Introduction}
\label{sec:introduction}
Acoustic perception remains underutilized in \glspl{av} and mobile robotics, although its physics complement those of optical sensors. Cameras and \acrshort{lidar} sensors are strictly line-of-sight and easily occluded, whereas sound diffracts around obstacles. An \gls{av} can ``hear'' an approaching emergency siren~\cite{yin2023realtime} or regular traffic~\cite{schulz2020hearing} from around a blind intersection long before the source comes into view, and a distributed microphone array provides this awareness all around the vehicle~\cite{marchegiani2021how}.

\begin{figure}[t]
    \centering
    \includegraphics{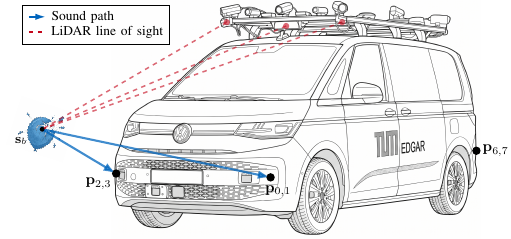} %
    \vspace{-8pt}\caption{Overview of the calibration setup. A balloon is burst at $\mathbf{s}_b$ near the vehicle. The impulse reaches the microphones (blue), giving $t_{b,m}^{\text{meas}}$, while the \acrshort{lidar} sensors (red) locate $\mathbf{s}_b$ itself.}\vspace{-8pt}
    \label{fig:photo_methodology}
\end{figure}

To fuse acoustic cues with the visual perception stack, the microphone array (\autoref{fig:photo_methodology}) must be metrically registered in the vehicle's 3D coordinate system, since an error in the array geometry appears directly in the estimated direction of arrival (\autoref{subsec:res-doa}). Established acoustic calibration methods assume small, unobstructed indoor rigs. A vehicle body, in contrast, casts an acoustic shadow, so sound reaching a microphone on the far side must diffract around the hull on a longer, \gls{nlos} path for which a straight-line propagation model has no term.

We propose a cross-modal calibration pipeline that uses the vehicle's existing cameras and \acrshortpl{lidar} to register the array in space and time. A balloon burst serves as a single multi-modal target. The \acrshort{lidar} locates the balloon, and thereby the source, in metric, scale-unambiguous coordinates, while the acoustic impulse provides a sharp onset at every microphone. The camera times the burst from the balloon's disappearance, which aligns the audio stream temporally with the other sensors. To cope with the acoustic shadow, the optimization weights each observation by its own measurement variance, which discounts the delayed, diffracted arrivals and achieves centimeter-level accuracy without prior knowledge of the microphone positions or the vehicle body.

\section{Related Work}
\label{sec:relatedwork}

\begin{table*}[t]
\centering
\setlength{\tabcolsep}{4pt}
\renewcommand{\arraystretch}{1.05}
\caption{State of the art in microphone array spatial calibration, ordered by proximity to this work. \emph{Aperture} is the largest microphone distance of the calibrated geometry or, where not reported, the extent of the experiment.}
    \label{tab:sota}
\footnotesize
\begin{tabularx}{\textwidth}{@{}l >{\raggedright\arraybackslash}p{2.6cm}
                              >{\raggedright\arraybackslash}X
                              >{\raggedright\arraybackslash}p{3.4cm}
                              >{\raggedright\arraybackslash}p{2.1cm}
                              >{\raggedright\arraybackslash}p{3.0cm} @{}}
\toprule
& \textbf{Work} & \textbf{Source (position from)} & \textbf{Aperture} &
\textbf{Position error} & \textbf{Evaluation} \\
\midrule

\multicolumn{6}{@{}l}{\emph{(a) Source localized by an independent second modality: the direct competitors}}\\
\addlinespace[1pt]
1 & Jacob and Haeb-Umbach~\cite{jacob2015absolute} & Moving human talker
(camera network, direction-of-arrival intersection) & Not reported (4 arrays in a
$6.2\times7.2$\,m room) & 20\,cm & \emph{Simulation only}
(image-method RIRs, modeled camera error) \\
2 & Li \textit{et al.}~\cite{li2025improved} & Six buzzers on a checkerboard
(camera) & 0.69\,m (0.4\,m cube) & 2.4--6.1\,cm &
Real hardware, indoor \\
\addlinespace[3pt]

\multicolumn{6}{@{}l}{\emph{(b) Source positions estimated jointly with the
geometry (\acrshort{slam})}}\\
\addlinespace[1pt]
3 & Zhang \textit{et al.}~\cite{zhang2024hybrid,zhang2025hybrid} & Chirps from a
robot-mounted loudspeaker (odometry, relative only) & Not reported (3 arrays
indoors, $1.6\times2\times1$\,m source trajectory) & 7.9\,cm &
Real hardware, indoor \\
4 & Wang \textit{et al.}~\cite{wang2024slam} & Chirps from a robot-mounted
loudspeaker (visual-inertial odometry, relative only) & Not reported (arrays spread over a
$15.5\times10$\,m area) & 23\,cm & Real hardware, indoor \\
\addlinespace[3pt]

\multicolumn{6}{@{}l}{\emph{(c) Passive: no controlled source}}\\
\addlinespace[1pt]
5 & Vanwynsberghe \textit{et al.}~\cite{vanwynsberghe2016robust} & Ambient
diffuse field (none) & 2.5 / $\sim$2\,m (128 / 64 mics) &
0.94 / 2.2\,cm & Real hardware, indoor \\
\midrule
$\bigstar$ & \textbf{Ours} & \textbf{Balloon burst (\acrshort{lidar})} &
\textbf{$\mathbf{4.9}$\,m (around the vehicle body)} & \textbf{$\mathbf{\ablRefHeldout}$\,cm}\textsuperscript{a} &
\textbf{Real hardware, outdoor, full vehicle} \\
\bottomrule
\end{tabularx}
\par\vspace{3pt}

\footnotesize
\textsuperscript{a}~Held-out range residual (\autoref{subsec:res-budget}); no sufficiently accurate external reference exists.
\end{table*}

\subsection{Microphone Array Spatial Calibration}
\label{subsec:rw-calibration}

Acoustic geometry calibration is surveyed by Plinge \textit{et al.}~\cite{plinge2016survey}, who note that acoustic measurements alone leave translation, rotation and possibly scale undetermined, so an external anchor is needed. \autoref{tab:sota} groups the prior work closest to ours by how the calibration source is located. Group~(a) localizes an active source with a second modality in its own known frame, the regime of this work. Group~(b) estimates source positions \emph{jointly} with the geometry in a \acrshort{slam} formulation, whose identifiability conditions~\cite{su2021observability,he2022observability} are only relaxed when source positions enter as measurements. Group~(c) forgoes a controlled source and calibrates from ambient or diffuse sound~\cite{hu2020passive,wozniak2024reverberation,vanwynsberghe2016robust} or from partial inter-microphone distances~\cite{taghizadeh2014adhoc}, and a related line estimates only an array's timing parameters~\cite{cao2023lowrank}.

Centimeter-level accuracy over a multi-meter aperture has been reached only indoors and only relative to the array, with a synchronized loudspeaker rig~\cite{plinge2016survey} or passively. Vanwynsberghe \textit{et al.}~\cite{vanwynsberghe2016robust} recover a $2.5$\,m array to under a centimeter, but only after an alignment step that absorbs any global-frame error. Joint estimation degrades sharply as arrays spread apart, from $0.20$\,m of position error at $1$\,m spacing to $2.56$\,m at $5$\,m~\cite{wang2024slam}.

\subsection{Limitations of Prior Work}
\label{subsec:rw-gaps}

\textit{1) No metric cross-modal reference.}
Only Jacob and Haeb-Umbach~\cite{jacob2015absolute} and Li \textit{et al.}~\cite{li2025improved} register the geometry to an external frame, both with a camera, whose pose is known only up to a scale that must be resolved separately and whose visual pattern is offset from the sources it locates. Elsewhere, a second modality fixes only the source's \emph{relative} motion~\cite{zhang2024hybrid,zhang2025hybrid,wang2024slam}. No prior work localizes the calibration source with a \acrshort{lidar}, which yields metric, scale-unambiguous positions directly in the vehicle frame.

\textit{2) No demonstration at vehicle scale outdoors.}
The largest real aperture with an independently measured source position is the $0.69$\,m diagonal of the $0.4$\,m cube of~\cite{li2025improved}, roughly seven times smaller than ours. The only geometry of comparable extent is simulated. Scale also changes which errors bind. Across $5$\,m, a sound-speed error equivalent to $3$\,K shifts the predicted arrival by some $2.6$\,cm and a $2\,\mathrm{m/s}$ wind component by a comparable amount, both negligible indoors, and a stochastic term grows with distance and turbulence~\cite{collier2004performance}. None of these appears in the error budgets of \autoref{tab:sota}.

\textit{3) Evaluation rarely reaches the downstream task.}
Most entries stop at parameter error. Vanwynsberghe \textit{et al.}~\cite{vanwynsberghe2016robust} propagate their geometry through a simulated beamformer, and Plinge \textit{et al.}~\cite{plinge2016survey} simulate how a speaker tracker degrades with calibration error, but neither measures the error its own procedure makes on the system it perturbs. Camera--\acrshort{lidar} robustness benchmarks~\cite{beemelmanns2024multicorrupt,yu2022benchmarking} also perturb extrinsics, but corrupt an already calibrated dataset to \emph{rank} architectures. A budget needs both halves on one system, the error the task tolerates and the error the procedure makes.

\textit{4) No scattering body in the propagation path.}
Every setup above is an unobstructed rig, for which straight-line propagation is a fair description. An array around a vehicle is not, since the hull shadows a large share of the source--microphone paths, and those arrivals are late by a deterministic excess the model has no term for. The survey~\cite{plinge2016survey} names such blocked paths as the outliers that undermine least-squares objectives, so far met with ad hoc countermeasures such as \gls{ransac}, and calls a principled treatment an open problem.

\subsection{Positioning and Contributions}
\label{subsec:rw-positioning}

We do not claim a new estimator, since batch nonlinear least squares on \gls{tdoa} measurements against known 3D source positions is an established technique~\cite{li2025improved,su2021observability}. A balloon burst is a broadband impulse with a sharp onset~\cite{patynen2011balloon}. The balloon is also the object the \acrshort{lidar} sees, so the measurement that locates it locates the acoustic source directly, without the marker-to-loudspeaker offset that a checkerboard-referenced rig must survey separately. Its radiation is two-lobed about the puncture point instead of omnidirectional~\cite{patynen2011balloon}, and randomizing the puncture direction across bursts (\autoref{sec:setup}) turns that systematic displacement of the acoustic origin into noise that averages down.

In summary, this work contributes:
\begin{itemize}
    \item \textbf{Metric cross-modal calibration:} registration of a microphone array in space and time from balloon bursts, with \acrshort{lidar} supplying metric 3D source positions.
    \item \textbf{Robustness to the acoustic shadow:} an observation weighting that discounts shadowed arrivals through their own variance, without a vehicle model, shown by ablation to outperform explicit masking.
    \item \textbf{Real-world evaluation:} full-vehicle hardware outdoors, with a downstream direction-of-arrival task that turns calibration error into an accuracy requirement.
\end{itemize}
\section{Method}
\label{sec:methods}

\begin{figure*}[t]
    \centering
    \resizebox{0.95\linewidth}{!}{\includegraphics{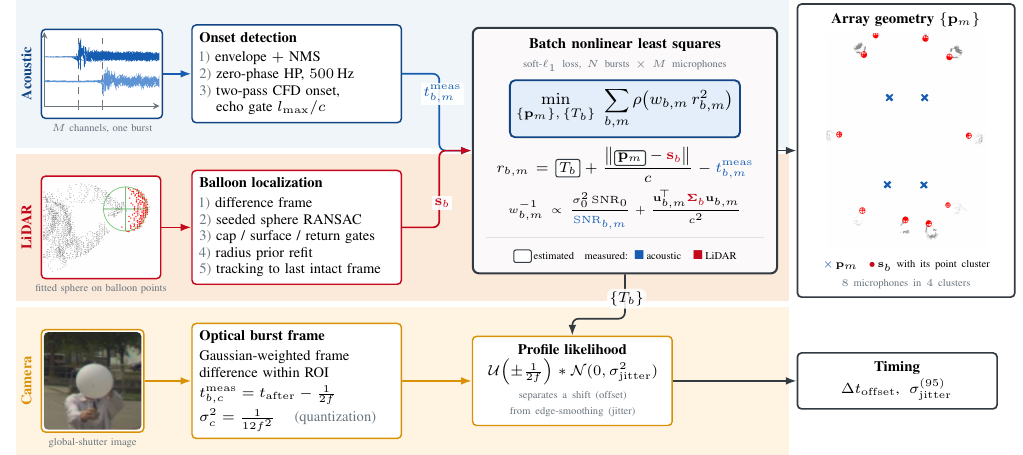}}%
    \vspace{-2mm}
    \caption{The calibration pipeline. Acoustic onset detection (top) and \acrshort{lidar} balloon localization (middle) supply the arrival times~$t^{\text{meas}}_{b,m}$ and source positions $\mathbf{s}_b$ for the batch nonlinear least-squares solve that recovers the array geometry $\{\mathbf{p}_m\}$ and burst times $\{T_b\}$ (right). A camera-referenced profile likelihood (bottom) separately calibrates the timing offset and jitter of the audio path.}\vspace{-4mm}
    \label{fig:methodology_pipeline}
\end{figure*}

\subsection{Problem Formulation}
\label{subsec:opt}
The pipeline of \autoref{fig:methodology_pipeline} recovers the microphone positions from the arrival times of balloon bursts at known source positions. Let $T_b$ be the unknown emission time of burst $b$ and $t_{b,m}^{\text{meas}}$ the measured arrival time of its wavefront at microphone $m$. With microphone position $\mathbf{p}_m$, source position $\mathbf{s}_b$ and speed of sound $c$, the predicted arrival time is
\begin{equation}
    \hat{t}_{b,m} = T_b + \frac{\|\mathbf{p}_m - \mathbf{s}_b\|}{c} ,
    \label{eq:forward}
\end{equation}
with the residual $r_{b,m} = \hat{t}_{b,m} - t_{b,m}^{\text{meas}}$. The calibration jointly estimates all microphone positions and burst times,
\begin{equation}
    \min_{\{\mathbf{p}_m\}, \{T_b\}} \sum_{b=1}^{N} \sum_{m=1}^{M} \rho\!\left( w_{b,m}\, r_{b,m}^{2} \right) ,
    \label{eq:optimization}
\end{equation}
where $w_{b,m}$ is the observation weight of \autoref{subsec:weighting} and $\rho(\cdot)$ a soft-$\ell_1$ loss, which bounds the influence of individual bad measurements. The problem is initialized by a coarse grid search over a common position of all microphones, and height bounds $0 \leq z_m \leq 2\,\mathrm{m}$ exclude the mirrored solution.

The speed of sound $c$ is computed per session from the measured air temperature, humidity and pressure~\cite{cramer1993speed} and held fixed instead of being estimated alongside the geometry. The reason is a near-degeneracy of \eqref{eq:forward}. For sources far compared with the aperture, the free emission time $T_b$ absorbs the common range, and what remains are the microphone offsets from the array centroid, projected onto the source direction and divided by $c$. Scaling all offsets and $c$ by the same factor therefore leaves \eqref{eq:forward} unchanged to first order in aperture over range, so array scale and sound speed form one direction that no residual criterion, cross-validation included, can separate. Fixing $c$ fixes the scale of the recovered geometry, and \autoref{subsec:res-ablation} shows what a free $c$ does instead.

\subsection{Acoustic Onset Detection}
\label{subsec:onset}

\begin{figure}[t]
    \centering
    \includegraphics{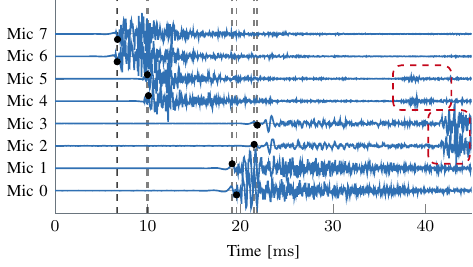} %
    \vspace{-5pt}\caption{Acoustic signals of a balloon burst. Black markers are the detected direct-path onsets $t_{b,m}^{\text{meas}}$, red boxes mark later reflections. On microphones 2 and 3, the reflection exceeds the attenuated direct wave in amplitude.}\vspace{-8pt}
    \label{fig:tdoa_example}
\end{figure}

Coarse burst times $t_b^{\text{meas}}$ are first found in the recording by threshold-based peak detection with non-maximum suppression on a global amplitude envelope over all channels. Around each burst, a window is cut from every channel and high-pass filtered at $500\,\mathrm{Hz}$, which removes wind noise and the low-frequency sound that blurs the onset. The filter runs forward and backward, so it adds no group delay, and its acausal pre-ringing shifts all onsets of a burst alike and is absorbed by the free emission time.

The onset in each channel is picked by a \gls{cfd}~\cite{gedcke1967cfd} with the threshold at $20\,\%$ of the largest absolute amplitude in the window, held at no less than five times the noise floor ahead of the burst. Since a reflection can exceed the direct wave in amplitude (\autoref{fig:tdoa_example}), the picker could lock onto an echo. A first pass therefore finds the earliest onset over all channels, which corresponds to the shortest path from the source. A bound $l_{\max}$ on the longest direct path, the array aperture plus the largest detour a wave diffracting around the vehicle can add, then confines the direct arrival at every microphone to a window of $l_{\max}/c$ after it, in which the picker is applied again. These onsets are the $t_{b,m}^{\text{meas}}$ of \eqref{eq:forward}.

\subsection{Balloon Localization in LiDAR}
\label{subsec:lidar}
\begin{figure}[b]
    \centering
    \subfloat[]{\includegraphics[width=0.485\linewidth]{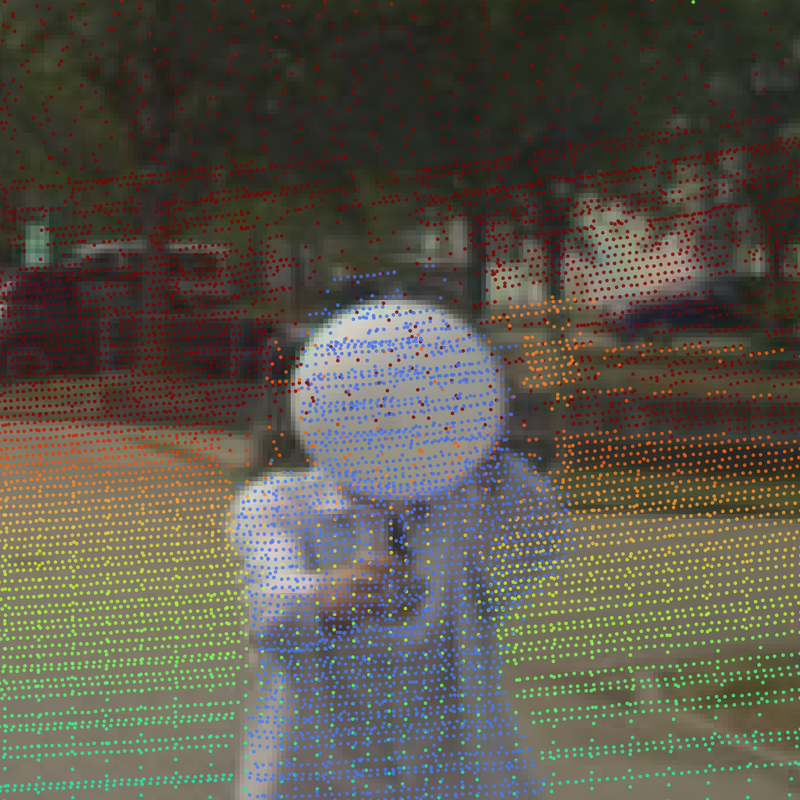}}
    \hfill
    \subfloat[]{\includegraphics[width=0.485\linewidth]{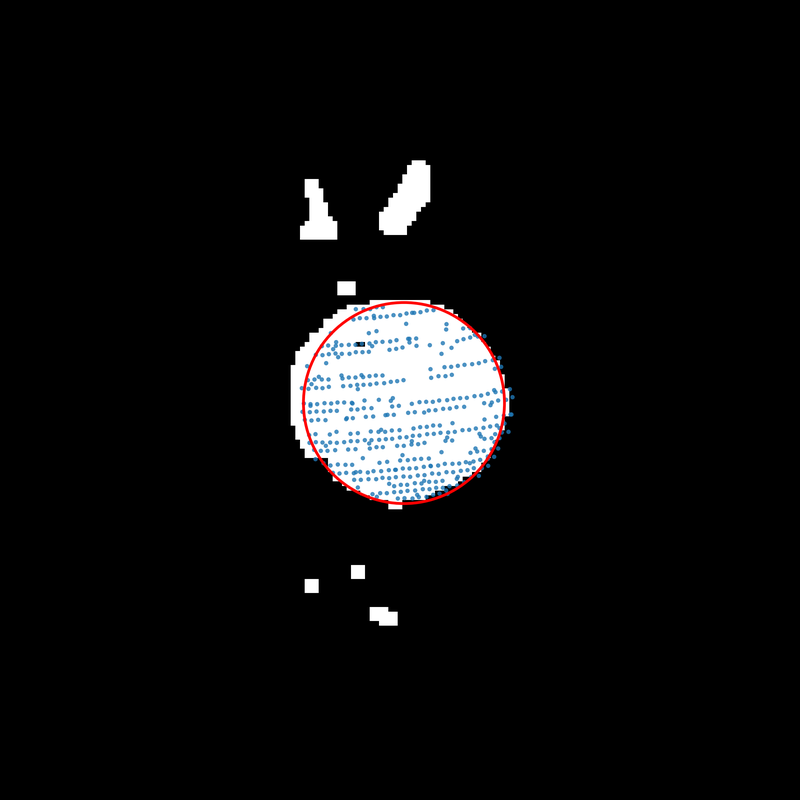}}%
    \caption{Balloon detection projected into the camera image. (a) Frame immediately before the burst, with the \acrshort{lidar} point cloud colored by range. (b) Binarized difference of the camera frames before and after the burst. Blue marks the \acrshort{lidar} difference-frame points, the red circle the fitted sphere of center $\mathbf{s}_b$ and radius $r$.}\vspace{-8pt}
    \label{fig:balloon_detection}
\end{figure}

We place the acoustic origin of the burst at the center of the sphere that best fits the balloon's rounded cap. Two \acrshort{lidar} frames about $0.2\,\mathrm{s}$ before and after the onset bracket the burst. The frame difference keeps every point of the earlier frame with no neighbor within $15\,\mathrm{cm}$ in the later one, i.e., the balloon and whatever else moved (\autoref{fig:balloon_detection}).

The difference frame is searched for a sphere of balloon-like radius. Around seed points on a coarse voxel grid, at three neighborhood radii, a sphere of center $\mathbf{s}$ and radius $r$ is fitted by \gls{ransac}~\cite{fischler1981ransac}, after an algebraic fit has discarded neighborhoods of implausible radius or residual. A candidate is kept only if it behaves like a burst balloon, that is, if the vanished points lie on a surface instead of filling a volume, have not returned in the later frame, and form a cap, not a band. Where the cap is too small to constrain $r$, the radius is fixed to the median of the well-covered fits.

Since a hand-held balloon can drift within the $\pm 0.2\,\mathrm{s}$ the anchor frames allow, the detected sphere is tracked forward frame by frame, counting the points left in its shell, until that count drops and does not recover. The sphere refitted in the last intact frame gives $\mathbf{s}_b$. Its center covariance $\mathbf{\Sigma}_b$, obtained by linearizing the sphere fit over a shell of raw points about that center with $r$ free, enters the weighting of \autoref{subsec:weighting}.

\subsection{Observation Weighting under Shadowing}
\label{subsec:weighting}
The forward model \eqref{eq:forward} assumes free-field propagation along the straight line between source and microphone. Around a vehicle, this does not hold for every observation. For any burst, some microphones lie on the opposite side of the body, so their direct path is blocked, and the acoustic energy arrives on an \acrshort{nlos} path diffracting around the hull. The detected onset then belongs to a path longer than the straight-line distance, and the residual $r_{b,m}$ would absorb this excess as if it were a position error. The excess is a deterministic, azimuth-dependent bias instead of noise, and amounts to about twice the held-out residual the calibration itself achieves, so it cannot be treated as a random error.

Excluding the shadowed pairs is not a viable alternative, because every burst then loses the far half of the array, which carries the longest baselines, and \autoref{subsec:res-ablation} shows this to be worse than keeping the biased observations. Instead, each residual is scaled by the inverse of its own arrival-time variance, to which two effects contribute. Onset detection degrades as a burst gets quieter, so the timing variance scales inversely with the observation's \gls{snr}, and the balloon-center uncertainty $\mathbf{\Sigma}_b$ enters through its component along the burst-to-microphone unit vector $\mathbf{u}_{b,m}$:
\begin{equation}
    w_{b,m} = \tilde{\sigma}^{2} \left(
        \sigma_{0}^{2}\frac{\mathrm{SNR}_{0}}{\mathrm{SNR}_{b,m}}
        + \frac{\mathbf{u}_{b,m}^{\top}\mathbf{\Sigma}_{b}\mathbf{u}_{b,m}}{c^{2}}
    \right)^{-1} .
    \label{eq:weight}
\end{equation}
Here $\mathrm{SNR}_{b,m}$ is the ratio of the largest amplitude inside the direct-arrival window of \autoref{subsec:onset} to the noise floor ahead of the burst, and $\sigma_{0}^{2}$ the onset-timing variance at a reference ratio $\mathrm{SNR}_{0}$. $\sigma_{0}$ is fixed once from the reference-free closure measurement of \autoref{subsec:res-budget}. The second term converts the position variance projected onto the ray into a time variance through $c$, and the squared median $\tilde{\sigma}^{2}$ of the per-observation standard deviation keeps the residuals in units of time.

Diffraction around the body both lengthens the path and attenuates the signal, so the biased observations are also the quiet ones, and \eqref{eq:weight} discounts them without knowledge of the vehicle or a classification threshold. A weight changes how strongly an observation pulls on the solution but not the direction of the pull, so the remaining bias is left to the measurement design. Bursts distributed around the vehicle make the shadowed set rotate with the source azimuth, so the displacements it induces oppose one another (\autoref{subsec:res-count}).

\subsection{Temporal Calibration}
\label{subsec:temporal}
The cameras and \acrshortpl{lidar} in our setup are \gls{ptp} synchronized, whereas the microphones are not. Their driver stamps sample blocks against the \acrshort{ptp}-disciplined clock, so the audio stream carries an unknown publish latency on a time base that advances with the audio device's own clock. That latency is estimated relative to the camera, whose global shutter, unlike a mechanical \acrshort{lidar} sweep, captures the balloon's disappearance at a single instant.

The burst position $\mathbf{s}_b$ is projected into the camera with the best view of it to define a \gls{roi}, and the largest Gaussian-weighted difference between neighboring grayscale frames identifies the pair of frames bracketing the burst (\autoref{fig:balloon_detection}(b)). The burst is placed at the midpoint of the interval preceding the first frame without the balloon, $t_{b,c}^{\text{meas}} = t_{\text{after}} - 1/(2f)$ at frame rate $f$, which leaves a uniform quantization error of variance $\sigma_c^2 = 1/(12f^2)$~\cite{widrow2008quantization}.

The emission time $T_b$ from \eqref{eq:optimization} is mapped onto the system clock by an affine fit over the whole recording, since a rate error of a few parts per million accumulates to milliseconds over minutes, and related to the camera by the standard offset model~\cite{furgale2013calibration}
\begin{equation}
    t_{b,c}^{\text{meas}} = T_b + \Delta t_{\text{offset}} + \eta_{\text{jitter}} + \eta_c ,
\end{equation}
with the latency $\Delta t_{\text{offset}}$, assumed constant within a recording, the audio jitter $\eta_{\text{jitter}} \sim \mathcal{N}(0, \sigma_{\text{jitter}}^2)$ and the camera quantization $\eta_c$, so that $\sigma_{\text{total}}^2 = \sigma_c^2 + \sigma_{\text{jitter}}^2$. The uncertainty of $T_b$ lies almost two orders of magnitude below $\sigma_c$.

The offset shifts the residual distribution, whereas the jitter rounds off the sharp edges introduced by quantization. The residual distribution is therefore modeled as the uniform quantization $\mathcal{U}(-\tfrac{1}{2f}, \tfrac{1}{2f})$ convolved with the Gaussian jitter $\mathcal{N}(0, \sigma_{\text{jitter}}^2)$, and both parameters are estimated by maximizing its likelihood, with the recordings pooled once their individual offsets are found to agree. The jitter can only be given an upper bound, because a finite sample cannot distinguish a small jitter from none at all. Requiring the audio path to inflate the total by no more than a fraction $\epsilon$ gives $\sigma_{\text{jitter}}^{(95)} \leq \sigma_c\sqrt{(1+\epsilon)^2-1} \approx \sigma_c\sqrt{2\epsilon}$, and we adopt $\epsilon = \tempEpsilon\,\%$. We also report the resolution floor $\sigma^{\min}(N)$, obtained by simulating the null at our sample size. Since the spatial solve fits every $T_b$ freely, a constant audio latency is absorbed into those estimates and leaves the array geometry unaffected, so the two calibrations can be run in sequence, and the latency is an output of the geometric calibration rather than an input to it.

\section{Experimental Setup}
\label{sec:setup}

All recordings were made on a full-size van converted into an automated driving research platform (\autoref{fig:photo_methodology}). Eight microphones are mounted low on the body in pairs near the four corners, with an aperture of $4.9\,\mathrm{m}$. The eight channels are recorded on a single sound card at $48\,\mathrm{kHz}$ and share one sampling clock. The \acrshort{ptp}-synchronized \acrshortpl{lidar} and cameras (\autoref{subsec:temporal}) run at $\tempFrameRate\,\mathrm{Hz}$.

The calibration campaign comprises $\tempNSolved$ balloon bursts recorded outdoors on open ground in $\tempNBags$ sessions on separate days. A person standing radially outward from the vehicle holds each balloon so that no body part lies between it and the array, and bursts it by hand. The burst height is varied across balloons so that the source positions are not coplanar, which would render the calibration unobservable~\cite{he2022observability}. The puncture direction is randomized across bursts, for the reason given in \autoref{subsec:rw-positioning}. All $\tempNSolved$ bursts enter the reference solution.

\section{Results and Discussion}
\label{sec:results}

\subsection{Temporal Calibration}
\label{subsec:res-temporal}

The temporal calibration is evaluated on $\tempNBursts$ bursts over $\tempNBags$ sessions, seen by $\tempNCams$ cameras at a frame period of $\tempFrameMs\,\mathrm{ms}$, which alone contributes $\sigma_c = \tempSigmaC\,\mathrm{ms}$. The per-session offsets agree to $\tempBagDiffSigma\,\sigma$ and the two halves of a session to $\tempDriftSigma\,\sigma$, so the latency is constant over a session and the sessions are pooled, giving $\Delta t_{\text{offset}} = \tempOffset \pm \tempOffsetHalfWidth\,\mathrm{ms}$ ($95\,\%$ interval), with the audio time base lagging the camera.

\begin{figure}[t]
    \centering
    \includegraphics{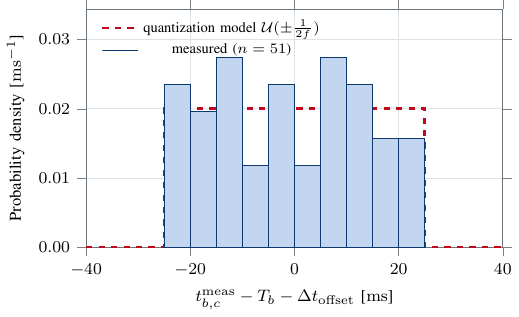} %
    \vspace{-18pt}\caption{Burst-time residual over $n = \tempNBursts$ bursts. The camera's frame quantization (dashed) sets the observed spread.}\vspace{-8pt}
    \label{fig:temporal_residual}
\end{figure}

The observed spread $\sigma_{\text{total}} = \tempSigmaTotal\,\mathrm{ms}$ is indistinguishable from the camera quantization, and the profile likelihood bounds $\sigma_{\text{jitter}}^{(95)} \leq \tempSigmaJitterBound\,\mathrm{ms}$, so the audio path inflates the total by $\tempInflation\,\%$, well inside the $\tempEpsilon\,\%$ allowed. That bound lies below the resolution floor $\sigma^{\min}(\tempNBursts) = \tempSigmaMin\,\mathrm{ms}$, so no jitter was resolved, rather than $\tempSigmaJitterBound\,\mathrm{ms}$ of it measured. The residuals in \autoref{fig:temporal_residual} fill the frame interval evenly and stop at its edges, with no bell-shaped component of their own. A sharper bound would not change the geometry, since the jitter it bounds is common to all channels of a burst and, like the latency itself, is absorbed by the free emission time.

\subsection{Error Budget}
\label{subsec:res-budget}

Two errors enter a predicted arrival, and both can be measured without a reference. Around a triangle of microphones the three pairwise \acrshort{tdoa}s sum to zero by construction, the zero cyclic sum condition of~\cite{scheuing2008disambiguation}, so the scatter of that sum is pure measurement noise. Since differences of per-channel onsets close identically, the test uses independent pairwise cross-correlations anchored at the onsets, whose noise stands in for the onset noise. Over all $\ablClosureTriangles$ triangles of the acoustically detected bursts, including those without a \acrshort{lidar} detection, it puts the per-arrival noise at $\ablClosureSigmaMs\,\mathrm{ms}$, or $\ablClosureSigmaCm\,\mathrm{cm}$ of path length. Propagating each burst's sphere-fit covariance onto its ray to a microphone adds $\ablRaySigmaPos\,\mathrm{cm}$. Together they predict a floor of $\ablPredictedFloor\,\mathrm{cm}$ for a calibration that has extracted everything the data holds, and the reference solution's cross-validated held-out residual is $\ablRefHeldout\,\mathrm{cm}$. The floor is set almost entirely by onset timing. The gap between floor and residual bounds what the floor omits: the displacement of the acoustic origin from the sphere center, at most one balloon radius, and any stochastic propagation delay, which cancels in the closure sum.

The only independent reference, a tape-measure survey of the microphones, agrees with the solution to within a decimeter once a known \acrshort{lidar}-to-vehicle translation is removed, which rules out gross errors but cannot certify a result at the noise floor, so the calibration is validated through the downstream task in \autoref{subsec:res-doa}.

\subsection{Ablation of the Shadow Handling}
\label{subsec:res-ablation}

\begin{table*}[t]
\centering
\caption{Ablations of the propagation handling. $\bar r$ is the \acrshort{rms} range residual, predicted minus measured path length, \emph{held-out} on balloons excluded from the fit and \emph{in-sample} split by whether the hull blocks the ray. $\Delta$ is its change against the full method, $\pm$ the standard error over $\ablFolds$ folds. $\Delta p$ is the median microphone displacement from the reference solution and \emph{Outl.} the number of balloons whose \acrshort{rms} range residual exceeds $25\,\mathrm{cm}$.}
    \label{tab:ablations}
\footnotesize
\newcolumntype{R}[1]{>{\raggedleft\arraybackslash}p{#1}}
\begin{tabularx}{0.92\textwidth}{@{}>{\raggedright\arraybackslash}X R{1.4cm} R{2.0cm} R{1.1cm} R{1.1cm} R{2.4cm} R{0.9cm}@{}}
\toprule
& \multicolumn{4}{c}{\textbf{\acrshort{rms} range residual} $\bar r$ \textbf{[cm]}} &
\textbf{Microphone} & \\
\cmidrule(lr){2-5}
\textbf{Variant} & \textbf{Held-out} & \textbf{$\Delta$ vs.\ ref.} &
\multicolumn{2}{c}{\textbf{In-sample}} &
\textbf{displacement $\Delta p$} & \textbf{Outl.} \\
& \footnotesize\itshape \acrshort{los} rays & \footnotesize\itshape $\pm$ SE &
\footnotesize\itshape \acrshort{los} & \footnotesize\itshape \acrshort{nlos} &
\footnotesize [cm] & \\
\midrule
\textbf{Full method (reference)}
  & \ablRefHeldout & --- & \ablRefInLos & \ablRefInNlos & \ablRefShift & \ablRefOutliers \\
\addlinespace[3pt]
\multicolumn{7}{@{}l}{\itshape Observation weighting} \\
\quad Without the source-position term
  & \ablNoCovHeldout & $\ablNoCovDelta \pm \ablNoCovSE$
  & \ablNoCovInLos & \ablNoCovInNlos & \ablNoCovShift & \ablNoCovOutliers \\
\quad Without the \acrshort{snr} term
  & \ablNoSnrHeldout & $\ablNoSnrDelta \pm \ablNoSnrSE$
  & \ablNoSnrInLos & \ablNoSnrInNlos & \ablNoSnrShift & \ablNoSnrOutliers \\
\quad No weighting at all
  & \ablNoWeightHeldout & $\ablNoWeightDelta \pm \ablNoWeightSE$
  & \ablNoWeightInLos & \ablNoWeightInNlos & \ablNoWeightShift & \ablNoWeightOutliers \\
\addlinespace[3pt]
\multicolumn{7}{@{}l}{\itshape Explicit shadow handling} \\
\quad Nearest-$\ablTrustedMics$ mask, others at weight $0.3$
  & \ablMaskPartialHeldout & $\ablMaskPartialDelta \pm \ablMaskPartialSE$
  & \ablMaskPartialInLos & \ablMaskPartialInNlos & \ablMaskPartialShift & \ablMaskPartialOutliers \\
\quad Nearest-$\ablTrustedMics$ mask, others excluded
  & \ablMaskFullHeldout & $\ablMaskFullDelta \pm \ablMaskFullSE$
  & \ablMaskFullInLos & \ablMaskFullInNlos & \ablMaskFullShift & \ablMaskFullOutliers \\
\quad One-sided loss, $\tau = 0.9$
  & \ablOneSidedHeldout & $\ablOneSidedDelta \pm \ablOneSidedSE$
  & \ablOneSidedInLos & \ablOneSidedInNlos & \ablOneSidedShift & \ablOneSidedOutliers \\
\addlinespace[3pt]
\multicolumn{7}{@{}l}{\itshape Sound speed} \\
\quad $c$ estimated jointly
  & \ablFreeCHeldout & $\ablFreeCDelta \pm \ablFreeCSE$
  & \ablFreeCInLos & \ablFreeCInNlos & \ablFreeCShift & \ablFreeCOutliers \\
\bottomrule
\end{tabularx}
\end{table*}

\autoref{tab:ablations} isolates the choices that address the propagation problem of \autoref{subsec:weighting}. For scoring, a ray counts as shadowed (\acrshort{nlos}) when its straight segment from burst to microphone intersects the \acrshort{cad} model of the vehicle, and as \gls{los} otherwise. That marks $\ablBlockedRays$ of the campaign's $\ablTotalRays$ rays, whose arrivals are late by $\ablNlosExcess\,\mathrm{cm}$ of excess path on average, against a \acrshort{los} population centered on zero. Every row changes one choice, is re-solved and is scored on the same folds, reference positions and $\ablClearRays$ \acrshort{los} rays. A row differs from the full method when $|\Delta|$ clearly exceeds its standard error.

No explicit model of the vehicle enters the optimization. The observation weighting handles the shadowed rays implicitly, and removing it costs $\unsigned{\ablNoWeightDelta} \pm \ablNoWeightSE\,\mathrm{cm}$ of held-out residual, effectively all of it through the \acrshort{snr} term. The source position is not the limiting error on this hardware, $\ablSigmaPos\,\mathrm{cm}$ of standard deviation along the ray against $\ablClosureSigmaCm\,\mathrm{cm}$ from onset timing, so its term is kept on principle.

Handling the shadow explicitly makes the calibration worse, monotonically. A diffraction model, as used for localization around known obstacles~\cite{an2019diffraction}, would need a hull accurate to the centimeter, which the available \acrshort{cad} model is not, so the alternatives tested are model-free. The mask trusts the~$\ablTrustedMics$ microphones nearest to each burst, a model-free rule that agrees with the hull test on $\ablMaskAgreement\,\%$ of the rays. Down-weighting the remaining rays to $0.3$ costs $\unsigned{\ablMaskPartialDelta}\,\mathrm{cm}$, and excluding them costs $\unsigned{\ablMaskFullDelta}\,\mathrm{cm}$, displaces the microphones by $\ablMaskFullShift\,\mathrm{cm}$ (median) and raises the outliers from $\ablRefOutliers$ to $\ablMaskFullOutliers$. Those rays see the array from the far side, and that geometry outweighs their bias. None of this shows in the fit's residual on the trusted rays, which \emph{falls} from $\ablDistMaskInLosRef$ to $\ablDistMaskInLosFull\,\mathrm{cm}$ as the held-out residual rises from $\ablRefHeldout$ to $\ablMaskFullHeldout\,\mathrm{cm}$.

A loss asymmetric in the residual sign, motivated by diffraction only ever lengthening a path, achieves nothing on the scored rays, $\unsigned{\ablOneSidedDelta} \pm \ablOneSidedSE\,\mathrm{cm}$, since the free emission time has already centered their residuals, yet it displaces the microphones by $\ablOneSidedShift\,\mathrm{cm}$ (median).

The last row shows why $c$ is held fixed. Estimated jointly with the geometry, it yields the table's best scores, as the held-out residual falls to $\ablFreeCHeldout\,\mathrm{cm}$ and the outlier count from $\ablRefOutliers$ to $\ablFreeCOutliers$, yet the result is physically wrong. The fitted $\ablFreeCSpeed\,\mathrm{m/s}$ lies far outside the $\ablMeasuredCLo$--$\ablMeasuredCHi\,\mathrm{m/s}$ of the recorded conditions, corresponds to an air temperature of $\ablFreeCTemp\,^\circ\mathrm{C}$, and comes with an array displaced by $\ablFreeCShift\,\mathrm{cm}$. This is the degeneracy of \eqref{eq:forward} made visible, since scale and sound speed trade against each other and cross-validation applies the same scaled pair to the held-out sources. As under the asymmetric loss, the failure is invisible to any residual statistic and is caught only against an external reference.

\subsection{Number and Placement of Bursts}
\label{subsec:res-count}

\begin{figure}[t]
    \centering
    \includegraphics{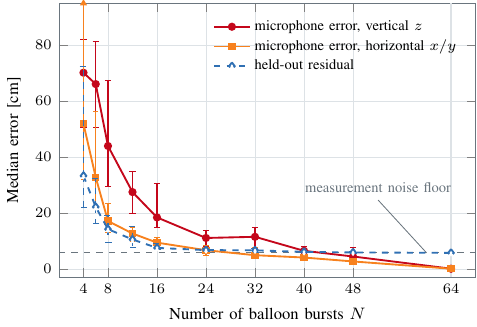} %
    \vspace{-3pt}\caption{Calibration quality against the number of bursts, over $\cntDraws$ draws per $N$. Markers are medians, bars the interquartile range. The horizontal error is per coordinate, comparable to the vertical one. Microphone error is measured against the $\cntRefN$-burst solution and vanishes there by construction, while the held-out residual settles at the noise floor of the fixed evaluation set (dashed).}\vspace{-8pt}
    \label{fig:count_sweep}
\end{figure}

How many bursts to record, and where, is answered by re-solving on subsets of the campaign. A fixed subset of $\cntEvalN$ balloons is withheld from every fit and every subset is scored on it, leaving $\cntRefN$ bursts to draw from, and microphone displacement is measured against the solution that uses all of them. \autoref{fig:count_sweep} shows both over $\cntDraws$ random draws per count. The displacement converges to zero because the subset grows into the reference, while the residual settles at $\cntFloor\,\mathrm{cm}$, the noise floor of this particular evaluation set. The held-out residual depends on which balloons are held out, ranging from $\ablFoldLo$ to $\ablFoldHi\,\mathrm{cm}$ over the folds of \autoref{subsec:res-budget}, and no number of balloons lowers it. From $N = 16$ to $N = 32$ the residual falls only from $\cntResOneSix$ to $\cntResThreeTwo\,\mathrm{cm}$ while the horizontal displacement halves, from $\cntHorizOneSix$ to $\cntHorizThreeTwo\,\mathrm{cm}$. Twelve well-spread bursts suffice where the residual is what matters, whereas an application that needs the geometry itself is repaid by twice as many.

The geometry does not improve evenly. The vertical error is about twice the horizontal one at every count and converges last, $\cntVertOneSix\,\mathrm{cm}$ at $N = 16$. This is a property of the campaign rather than of the estimator, because the bursts occupy a narrow elevation band, $\cntElevLo^\circ$ to $\cntElevHi^\circ$ with a median of $\cntElevMed^\circ$, seen from a nearly planar array, so barely a sixth of a vertical displacement is observable. The horizontal bearings this work targets are little affected. An application that needs elevation should widen that band instead of the number of bursts.

\begin{figure}[t]
    \centering
    \vspace{-8pt}\includegraphics{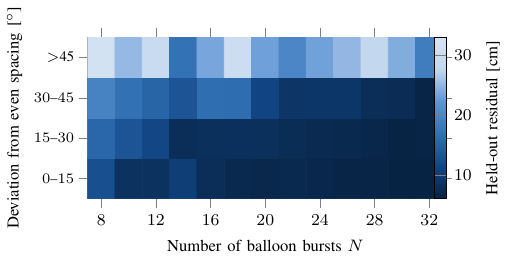} %
    \vspace{-16pt}\caption{Held-out residual against burst count and departure from even spacing in azimuth, over $\cntPlaceRuns$ constructed arrangements. Each cell is the median of at least $\cntPlaceMinCell$ arrangements.}\vspace{-8pt}
    \label{fig:placement_heatmap}
\end{figure}

Placement is varied separately. Arrangements of a given count are built with a controlled departure from even azimuth spacing, the \gls{rms} deviation of the realized bearings from the best-matching equally spaced set, and each is re-solved. \autoref{fig:placement_heatmap} summarizes $\cntPlaceRuns$ of them. At the low end, evenness is worth as much as count, as $\cntPlaceLo$ evenly spread bursts reach $\cntEvenEight\,\mathrm{cm}$ and beat the $\cntUnevenThreeTwo\,\mathrm{cm}$ of $\cntPlaceHi$ badly spread ones. The top row barely improves from left to right, so additional bursts do not buy back coverage, and time is better spent walking around the vehicle. The count study agrees. Drawing the subsets balanced over azimuth instead of at random improves $N = 12$ from $\cntResOneTwo$ to $\cntBalancedOneTwo\,\mathrm{cm}$, close to the $\cntResOneSix\,\mathrm{cm}$ of sixteen random ones, but leaves $N = 24$ unchanged, since by then the vehicle is covered anyway.

\subsection{Downstream Validation on a Moving Source}
\label{subsec:res-doa}

The array's internal geometry can be compared with a tape survey (\autoref{subsec:res-budget}), but its orientation in the \acrshort{lidar} frame cannot, because that frame's origin is an optical center inside a sensor housing. A direction-of-arrival task reaches exactly that half of the calibration. We therefore recorded a loudspeaker carried on foot around the stationary vehicle, playing a maximum-length sequence band-limited to $400$--$3000\,\mathrm{Hz}$, while the vehicle's perception stack tracked its carrier as a pedestrian. The tracked box gives the reference azimuth and, through its dimensions and orientation, the angle the source itself occupies. Over the $\pbDuration\,\mathrm{s}$ in which the sequence is audible, the source moves between $\pbRangeLo$ and $\pbRangeHi\,\mathrm{m}$ and sweeps from $\pbAzLo^\circ$ to $\pbAzHi^\circ$.

\begin{figure}[t]
    \centering
    \includegraphics{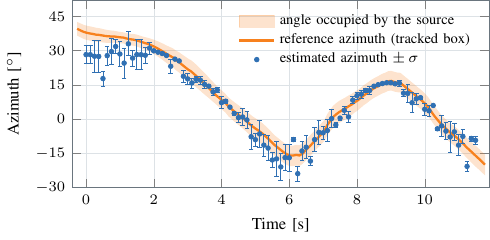} %
    \vspace{-10pt}\caption{Direction of arrival on a moving source. Estimated azimuth against the reference from the vehicle's own \acrshort{lidar} tracking, over $\pbN$ accepted windows of $\pbWindowMs\,\mathrm{ms}$. The band is the angle the source itself occupies, median $\pbExtent^\circ$, and error bars are the spread of the $\pbBands$ sub-band estimates.}\vspace{-8pt}
    \label{fig:azimuth_validation}
\end{figure}

The azimuth is estimated by near-field steered response power, which sums the pairwise \gls{gcc} with phase transform weighting over candidate source positions~\cite{grinstein2024srp}. The near-field form is needed since microphone separations are comparable to the source distance, and searching over positions instead of directions refers the reported azimuth to the origin of the \acrshort{lidar} frame, where the reference is defined. The estimate uses the four front microphones. Unlike the calibration, which could absorb the shadowed rays because the balloons surrounded the vehicle, the passing source lies on one side only, and the diffracted arrivals at the far-side microphones carry no consistent timing. \acrshort{tdoa} triangles within the front sub-array close to within $2\,\mathrm{ms}$ in $\pbClosureFront\,\%$ of the windows, whereas triangles spanning front and rear do so in only $\pbClosureMixed\,\%$. Each estimate uses $\pbWindowMs\,\mathrm{ms}$ of audio, over which the source moves less than the error measured, and is the median of $\pbBands$ sub-band estimates, whose spread serves as its uncertainty. Windows in which the signal falls more than $\pbGateDb\,\mathrm{dB}$ below its peak, after the loudspeaker was switched off, are discarded.

\begin{figure}[b]
    \centering
    \includegraphics{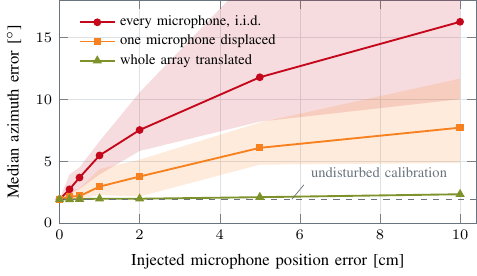} %
    \vspace{-8pt}\caption{The same windows are re-solved with the microphone geometry perturbed, one mode at a time, $\dgDraws$ draws per level, showing a rising error with higher position error. Lines are medians, bands the interquartile range.}\vspace{-8pt}
    \label{fig:azimuth_degradation}
\end{figure}

\autoref{fig:azimuth_validation} shows the result over $\pbN$ windows, with a median error of $\pbMedErr^\circ$, quartiles $\pbQOne$ and $\pbQThree^\circ$, $\pbWithinTen\,\%$ of estimates within $10^\circ$, and a mean offset of $\pbBias \pm \pbBiasSE^\circ$ over $\pbNeff$ effectively independent windows, at the edge of significance and consistent with the hand-carried loudspeaker moving off the center of the tracked body. That residual is comparable to the $\pbExtent^\circ$ the source itself subtends, so what remains is largely which part of the source radiates. Agreement at the scale of the object's own angular size is what associating an acoustic direction with a tracked object requires.

That this depends on the calibration, and not only on the estimator, is shown by re-solving the same windows with the microphone geometry perturbed (\autoref{fig:azimuth_degradation}). The error grows monotonically with every mode that changes the array's shape, to $\dgJitterOne^\circ$ for one centimeter of independent error per microphone and to $\dgSingleOne^\circ$ when a single microphone carries it. Translating the array as a whole is almost free, $\dgTransTen^\circ$ at ten centimeters, as it must be for a source outside the aperture, whereas a rotation of the array about the \acrshort{lidar} origin rotates the estimated azimuth with it, which lets this experiment reach the one component no other measurement here can.

\section{Conclusion and Outlook}
\label{sec:conclusion}

We presented a calibration of a vehicle-mounted microphone array that needs no prior knowledge of the microphone positions or the vehicle body. Balloon bursts serve as the target, located by \acrshort{lidar}, timed by their acoustic onset and by their optical disappearance, so that a single event registers the array in space and in time. The held-out range residual of $\ablRefHeldout\,\mathrm{cm}$ approaches the $\ablPredictedFloor\,\mathrm{cm}$ floor that the measured onset noise predicts, so what limits the result is acoustic onset timing, not the \acrshort{lidar}.

On a moving source the calibration places a passing loudspeaker to a median of $\pbMedErr^\circ$, accurate enough to associate an acoustic direction with a tracked object, and degrades to $\dgJitterOne^\circ$ under one centimeter of per-microphone error. Discounting the shadowed arrivals by their own variance, instead of masking them, keeps the array's longest baselines in the fit, and no model-free mask tested did better. For the field, twelve evenly spread bursts suffice, and spreading them buys more than adding more.

Two questions stay open. An independent survey of the microphone positions would turn the held-out residual into a true accuracy figure and reveal any bias that the shadowed rays impose and that scoring on unobstructed rays cannot see. A diffraction model of the hull, if a \acrshort{cad} model accurate to the centimeter is available, could recover the shadowed arrivals instead of discounting them. The same bursts may also serve the measurement of a \gls{vrtf}, which requires this geometry.

  \section*{Acknowledgment}
  \noindent This work was partly funded by the German Federal Ministry of Research, Technology and Space (BMFTR) within the project ASUR under grant number 03ZU2105BA. The responsibility for the content lies with the authors. 
  
  \noindent The authors used Gemini (Google) to stylize the vehicle image in \autoref{fig:photo_methodology} and Claude (Anthropic) for coding and language editing, and reviewed all generated content.

\bibliographystyle{IEEEtran}
\bibliography{literature}
\end{document}